\documentstyle[12pt]{article}
\title{
 Chiral-symmetry restoration in the linear sigma model at
nonzero temperature and baryon density
 }
\author{
Neven Bili\'c
and
Hrvoje Nikoli\'c  \\
Rudjer Bo\v{s}kovi\'{c} Institute, \\
P.O. Box 1016, 10001 Zagreb, Croatia \\
\makebox[1in]{} \\
 IRB-TP 229/97,
 \hspace{0.1in}
 hep-ph/9711513
}
\date{\today}
\begin{document}
\maketitle
\begin{abstract}
We study the chiral phase transition in the linear
sigma model with 2 quark flavors and
$N_c$ colors.
 One-loop calculations predict a
first-order phase transition at both
$\mu=0$ and $\mu\neq 0$.
We also discuss the phase diagram
 and make a comparison with a
 thermal parametrization of
existing heavy-ion experimental data.
\end{abstract}

%

\section{Introduction}

The linear sigma model,
originally proposed as a  model for
strong nuclear interactions
\cite{gel},
today serves as an effective
model for the low-energy (low-temperature)
phase of quantum chromodynamics.
The model exhibits spontaneous breaking
of chiral symmetry
 and restoration
at finite temperature.
In this paper we discuss the pattern of symmetry breaking
and its restoration at nonzero temperature
and nonzero chemical
potential at one-loop level.
Some aspects of this  have
been extensively studied in the literature
[2-12].
The phenomenological importance
of the chiral transition and
possible experimental signatures
have recently been discussed by Harris and M\"uller
\cite{har}.
In connection with  theoretical predictions of the phase transition
 there still exist a certain misunderstanding and controversy.
In particular, the precise value of the critical
temperature is not generally agreed on
\cite{ram,cle,con,lur,bil}.
As regards the order of the transition,
Pisarski and Wilczek \cite{pis} have shown on  general grounds
that chiral models with 2 massless flavors
undergo a first-order phase transition at finite
temperature.
In actual calculations, the mean field
 predicts a second-order transition \cite{bay,boc}, whereas
some improved mean-field results indicate a weak
first-order transition \cite{lar}.
Numerical simulation \cite{gau} of a three-dimensional
chiral model on the lattice also confirms a
first-order transition for $N_f=2$.
 We believe that in our approach we are able
to clear up some of these points.
In addition, we discuss
the phase diagram and its relation
to  heavy-ion collisions.

We organize the paper as follows.
In Sect. \ref{eff} we describe the thermodynamics
of the model.
In Sect. \ref{pimass}
 we calculate the temperature
dependence of the pion and sigma masses.
In Sect. \ref{chiral} we discuss
the chiral phase transition
and make
comments on and a comparison  with other papers.
 In the concluding section,
Sect. \ref{concl}, we summarize our results.

\section{Effective potential}  \label{eff}

The linear sigma model of Gell-Mann and L\'evy
is an effective model of strong interactions
described by the chirally symmetric Lagrangian (in Euclidean notation)
\begin{eqnarray}
\label{eq1}
{\cal{L}}
         & = &
\bar{\psi} (\gamma_{\mu} \partial_{\mu} +
g (\sigma + i \mbox{\boldmath $\tau \pi$} \gamma_5)) \psi -\mu
\bar{\psi} \gamma_4 \psi
 \nonumber  \\
 &   &
+ {1\over 2} (\partial \sigma)^{2}
+{1\over 2} (\partial
\mbox{\boldmath $\pi$})^{2}
 + {m_0^{2}\over 2} (\sigma^{2} +
\mbox{\boldmath$\pi$}^{2})
+ {\lambda\over 4} (\sigma^{2} +
\mbox{\boldmath$\pi$}^{2})^{2}
\; .
\end{eqnarray}
The right and left fermions
$\psi_R=\frac{1}{2} (1+\gamma_5)\psi$,
$\psi_L=\frac{1}{2} (1-\gamma_5)\psi$,
 are assumed to constitute, respectively, the
$ (\frac{1}{2},0)$ and $(0,\frac{1}{2})$
 representation of the chiral SU(2)$\times$SU(2),
 whereas the mesons ($\sigma$,
{\boldmath$\pi$}) belong to
the $(\frac{1}{2},\frac{1}{2})$ representation.
In the original sigma model \cite{gel} the fermion field
was a nucleon.
We consider
 the  fermions to be
constituent quarks \cite{con,goc} with an
additional degree of freedom,
``color'', from the SU($N_{c}$) local gauge group of
an underlying gauge theory (QCD).

 If $m_0^{2} < 0$, the chiral
symmetry will be spontaneously broken.
At the classical level, the $\sigma$ and $\pi$ fields develop
nonvanishing expectation values such that
\begin{equation}\label{eq2}
\langle \sigma \rangle^{2} + \langle \mbox{\boldmath$\pi$} \rangle^{2}=
- {m_0^{2}\over \lambda} \equiv f_{\pi}^{2} \; .
\end{equation}
It is convenient to choose here 
\begin{equation}\label{eq3}
\langle \pi_{i} \rangle = 0, \;\;\;\;\; \; \langle \sigma \rangle =
f_{\pi} \; .
\end{equation}

In order to study the thermodynamics of the model, we define the
thermodynamical potential as a function of the chemical potential $\mu$
associated with baryon number density and inverse
temperature $\beta = 1/T$
\begin{equation}\label{eq7}
\Omega (\beta, \mu) = - {1\over \beta V} \ln \; Z \;,
\end{equation}
where the partition function $Z$ is defined as a path integral
\begin{equation}\label{eq8}
Z = \int \; [d \varphi] \; \; \exp \{-\int_{0}^{\beta} d \tau \int
d^{3}x \; {\cal{L}} (\varphi)\} \; .
\end{equation}
Here [$d \varphi$] is an abbreviation for
 the integral over $\psi$, $\sigma$,
{\boldmath$\pi$}, and ${\cal{L}}$ is given by (\ref{eq1}).
Next we introduce
the saddle-point method of Frei and Patk\'{o}s
 \cite{fre}.
Our approach is similar to that of Meyer-Ortmanns
and Schaefer \cite{mey} who applied the method
to the chiral SU(3)$\times$SU(3.)
 We first redefine the fields
\begin{eqnarray}\label{eq9}
\mbox{\boldmath$\pi$} & \rightarrow &
\mbox{\boldmath$\pi$} + \mbox{\boldmath$\pi$}'(x) \; ,
\nonumber \\
\sigma & \rightarrow & \sigma + \sigma'(x) \; ,
\end{eqnarray}
where {\boldmath$\pi'$} and $\sigma'$
are quantum fluctuations around the
constant values {\boldmath$\pi$} and $\sigma$,
respectively.
Next we use the transformation which
 quadratizes the quartic interaction
\begin{equation}\label{eq100}
 \exp \{-\int d^4x \, \frac{\lambda}{4}
 (\sigma'^2 +
\mbox{\boldmath $\pi$}'^2)^2 \} =
\int_{c-i\infty}^{c+i\infty} [ds] \,
 \exp \{-\int d^4x \,
[ \frac{s}{2}(\sigma'^2 +
\mbox{\boldmath $\pi$}'^2)
 -\frac{s^2}{4\lambda}] \}.
\end{equation}
and redefine the auxiliary field
\begin{equation}\label{eq140}
s(x) \rightarrow s+s'(x) \; ,
\end{equation}
so that the saddle-point value $s$ maximizes
the integrand.
     The thermodynamical
potential $\Omega$ as a function of $\sigma$, {\boldmath$\pi$}
and $s$ is
usually called the {\em effective potential}.
 Thermodynamics requires that, in
thermal equilibrium, $\Omega$ should assume a minimum with respect to
variations of {\boldmath$\pi$} and $\sigma$.
 Owing to the specific
form of the interaction and
because of (\ref{eq2}) and (\ref{eq3}),
 we can keep $\mbox{\boldmath$\pi$} = 0$ and
consider $\Omega$ as a function  of
the two mean fields $\sigma$ and $s$.
 Using (\ref{eq9}), (\ref{eq100}) and (\ref{eq140})
 we find
\begin{equation}\label{eq10}
\Omega(\sigma,s) =
 {\lambda\over 4} \; \sigma^{4} + {m_0^{2}\over
2} \; \sigma^{2}-{s^2 \over 4\lambda}
-{1\over \beta V} \; \ln \; Z'(\sigma,s) \; ,
\end{equation}
where $Z'$ is the partition function for the shifted Lagrangian
 in which
 the quartic interaction is absent and
 chiral symmetry is explicitly broken:
\begin{eqnarray}\label{eq5}
{\cal{L}}' & = &
  \bar{\psi} (\gamma \partial + m_{F} + g
(\sigma' + i \; \mbox{\boldmath$\tau$}
\mbox{\boldmath$\pi$}' \gamma_{5})) \psi
- \mu \bar{\psi} \gamma_4 \psi
+{1\over 2} (\partial
\mbox{\boldmath $\pi$})'^{2}
 + {1\over 2} (\partial \sigma')^{2}
\nonumber  \\
  &  &
+ {m_{\sigma}^{2}\over 2} \sigma'^{2} + {m_{\pi}^{2}\over 2}
\mbox{\boldmath$\pi$}'^{2}
-{1 \over 4\lambda} s'^2
+g' \sigma' (\sigma'^2 +
\mbox{\boldmath $\pi$}'^{2})
 + {s'\over 2}
 (\sigma'^2 +
\mbox{\boldmath $\pi$}'^{2})
 + c  \sigma' -{1 \over 2\lambda}ss' \, .
\end{eqnarray}
 The effective masses,
  the trilinear coupling $g'$ and $c$ are functions
of $\sigma$ and $s$ defined as
\begin{eqnarray}\label{eq11}
m_{\sigma}^{2} & = &
 m_0^{2} + s+3 \lambda \sigma^2  \,  ,  \; \;\;\;
  m_{F} =g \sigma \, ,
  \nonumber \\
m_{\pi}^{2} & = & m_0^{2}+s+\lambda \sigma^{2} \, , \;\;\;\;
g' =  \lambda \sigma  \, ,
  \nonumber \\
c & = & \sigma(m_0^{2} + \lambda \sigma^{2}) \; .
\end{eqnarray}
The condition for an extremum
\begin{equation}\label{eq12}
{\partial \Omega\over \partial \sigma} = 0 \;, \;\;\;
{\partial \Omega\over \partial s} = 0 \; ,
\end{equation}
gives two equations for $\sigma$ and $s$
 with solutions that
will in general depend on temperature and
chemical potential.
A nontrivial solution $\sigma(\beta,\mu)$
 will be referred to as
{\em chiral condensate}.

At the classical level
(neglecting the quantum and thermal fluctuations)
 the potential
takes the  form
\begin{equation}\label{eq13}
\Omega(\sigma,s) = {1\over 4} \lambda  \sigma^{4} +
{m_0^{2}\over 2}  \sigma^{2}
-{s^2\over 4\lambda} \; .
\end{equation}
An extremum (minimum with respect to
$\sigma$, maximum with respect to $s$) occurs at
\begin{equation}\label{eq14}
\sigma^2 = -{m_0^{2}\over \lambda}
 \equiv f_{\pi}^{2} \; , \;\;\; s=0 \; ,
\end{equation}
yielding
\begin{equation}\label{eq15}
m_{\pi} = 0 \; , \; m_{F} = g f_{\pi} \; , \;
m_{\sigma}^2 =
2 \lambda f_{\pi}^2 \; , \; g' = \lambda f_{\pi} \;
, \; c'=0 \; .
\end{equation}
Thus, at the classical level, we
have 3 massless pions, as it should be owing to the Goldstone
theorem.

The thermal and quantum fluctuations
 will change the effective potential into
\begin{equation}\label{eq16}
\Omega (\sigma,s) = {\lambda\over 4} \sigma^{4} +
{m_0^{2}\over 2} \sigma^{2} -{s^2 \over 4\lambda}+
\Omega_{0} (\sigma,s) +
\Omega_{I} (\sigma,s) \; .
\end{equation}
 Here $\Omega_{I}$ contains
loop corrections and
$\Omega_{0}$ is the thermodynamical potential for a noninteracting
gas of  fermions
and bosons:
\begin{equation}\label{eq17}
\Omega_{0}  =  \Omega_{F} + \Omega_{\sigma} + \Omega_{\pi} \; ,
\end{equation}
\begin{eqnarray}\label{eq18}
\Omega_F & = &  -N_c N_f {1 \over \beta} \sum_l \int {d^{3}p \over
(2 \pi)^{3}} \; {\rm Tr} \; \ln
 \left[ \beta (- i {\not\!{p}}+ m_{F}) \right] \; ,
 \nonumber  \\
\Omega_{\sigma} & = & {1\over 2 \beta} \sum_n \int
{d^3 k\over (2 \pi)^3} \; \ln  \left[\beta^2 (k^2 + m_{\sigma}^2)
\right]   \; ,
 \nonumber  \\
\Omega_{\pi} & = & (N_f^2 - 1) \; {1\over 2 \beta} \sum_n \int
{d^{3} k\over (2 \pi)^{3}} \; \ln  \left[\beta^{2} (k^{2} + m_{\pi}^2)
\right] \; ,
\end{eqnarray}
where
\begin{equation}\label{eq19}
p = \left((2 l +1){\pi\over\beta}+i \mu \, ,
 \mbox{\boldmath $p$} \right)\; ,
\end{equation}
\begin{equation}\label{eq20}
k = \left(2 n {\pi\over\beta} \, , \mbox{\boldmath $k$} \right)\; ,
\end{equation}
and the number of flavors is $N_f=2$.
The extremum condition now reads
\begin{eqnarray}
{\partial \Omega\over \partial \sigma}
 & = &
  \lambda \sigma^3 + m_0^2
\sigma
+ {1\over \beta V} \int d^{4}x \: \biggl\{g \langle \bar{\psi}(x)
\psi(x) \rangle + 3 \lambda \sigma \langle \sigma'(x)^2 \rangle \biggr.
\nonumber  \\
&  &
+\lambda \sigma \langle \mbox{\boldmath$\pi$}'(x)^{2} \rangle
+ \lambda \langle \sigma'(x) (\sigma'(x)^{2} +
\mbox{\boldmath$\pi$}'(x)^{2}) \rangle
 + (m_0^{2} + 3 \lambda \sigma^{2}) \biggl. \langle
\sigma'(x) \rangle \biggr\} = 0 \, ,
\label{eq21} \\
{\partial \Omega\over \partial s}
&  = & -{s \over 2\lambda} +
 {1\over \beta V} \int d^{4}x \:
\biggl\{ {1 \over 2}\langle \sigma'(x)^{2} +
\mbox{\boldmath$\pi$}'(x)^{2} \rangle
-{1 \over 2\lambda}\langle s'(x) \rangle  \biggr\} =0 \; .
\label{eq210}
\end{eqnarray}
The terms $\langle \sigma' \rangle $ and
 $\langle s' \rangle $
 vanish  because the quantum fluctuations take place
around the true vacuum.
Equations (\ref{eq21}) and (\ref{eq210}) become
\begin{equation}\label{eq22}
\lambda \sigma^{3} + m_0^{2} \sigma - 2 g N_{c}
 {\cal{G}}_{F} + 3 \lambda \sigma [{\cal{G}}_{\sigma}
+ {\cal{G}}_{\pi}]
+\lambda [\Gamma_{\sigma\sigma\sigma}
+\Gamma_{\sigma\pi\pi}]=0 \, ,
\end{equation}
\begin{equation}\label{eq220}
-{s \over 2\lambda}+
{1 \over 2}{\cal{G}}_{\sigma}
+ {3 \over 2} {\cal{G}}_{\pi} = 0 \; .
\end{equation}
Here ${\cal{G}}$ and $\Gamma$
denote thermal averages (with respect to the {\bf full}
 partition function) over a product of two
 and three fields, respectively.
 For example,
\begin{equation}\label{eq23}
{\cal{G}}_{\sigma} = {1\over \beta V} \; \int \; d^4 x \;\langle
\sigma(x)^{2} \rangle \, .
\end{equation}
This coincides with the full 2-point Green's function at $x = 0$.
${\cal{G}}$ is often referred to as {\em tadpole}.
Equations (\ref{eq22}) and (\ref{eq220}), schematically depicted
in Fig. \ref{fig1},
state the fact that tadpoles cancel \cite{ger} also at
nonzero temperature and chemical potential.

\begin{figure}[p]
\caption{Schematic representation of (23) and (24)}
\label{fig1}
\end{figure}

Solutions to  (\ref{eq22})
and (\ref{eq220}) are implicit functions of $T$ and $\mu$.
The tadpoles
at one-loop order are given by
\begin{eqnarray}\label{eq25}
{\cal{G}}_{F} & = &
  {1 \over \beta} \sum_l \int {d^{3}p \over (2\pi)^3}
 \; {\rm Tr} \;
{1\over - i {\not\!{p}} + m_{F}}   \; ,
\nonumber  \\
{\cal{G}}_{\sigma, \pi} & = &
  {1 \over \beta} \sum_n \int {d^{3}k \over (2\pi)^3 }
{1\over k^{2} + m_{\sigma,\pi}^{2}} \; ,
\end{eqnarray}
whereas the three-point functions
$\Gamma_{\sigma\sigma\sigma}$ and
$\Gamma_{\sigma\pi\pi}$
 contribute at two-loop (and higher) order.
The masses in (\ref{eq25}) depend on
 $\sigma$ and $s$ through  (\ref{eq11}).
Equation (\ref{eq22}) has, apart from $\sigma
= 0$, a nontrivial solution
$\sigma(\beta,\mu)$ that no longer equals $f_{\pi}$.
Dividing  it by $\lambda \sigma$,
this equation at one-loop order may be written as
\begin{eqnarray}\label{eq250}
\sigma^2
&=&
 f_{\pi}^{2} + 4 \: {2 g^2 \over \lambda}
N_{c} \; {1\over \beta}\sum \int {d^{3} p\over (2\pi)^3}
\frac{1}{p^2+m_{F}^{2}}
\nonumber  \\
&  &
-  3 \: {1\over \beta}\sum \int {d^3 k\over (2 \pi)^3}
\; {1\over k^{2} + m_{\sigma}^{2}}
-  3 \: {1\over \beta}\sum \int {d^{3} k\over(2\pi)^3}
\frac{1}{k^{2} + m_{\pi}^{2}} \; ,
\end{eqnarray}
where we have replaced $-m_0^{2}/\lambda = f_{\pi}^{2}$ from
(\ref{eq14})
and used $m_{F} = g \sigma$.
Similarly, (\ref{eq220}) becomes
\begin{equation}\label{eq251}
s =
\lambda \: {1\over \beta}\sum \int {d^3 k\over (2 \pi)^3}
\; {1\over k^{2} + m_{\sigma}^{2}}
+ 3 \lambda \: {1\over \beta}\sum \int {d^{3} k\over(2\pi)^3}
\frac{1}{k^{2} + m_{\pi}^{2}} \; .
\end{equation}

We can separate the finite $T$ and $\mu$ part of ${\cal{G}}$ as usual
 \cite{kap}:
\begin{equation}\label{eq26}
{\cal{G}}_{F}  = 4\int {d^4 q\over (2 \pi)^{4}} \;
{m_F\over q^2 + m_{F}^{2}}
- 4 \int  {d^{3} q\over (2 \pi)^{3}} \; {m_{F}\over 2
\omega_{F}} \; n_{F} (\omega_{F}) \, ,
\end{equation}
and
\begin{equation}\label{eq27}
{\cal{G}}_{\sigma, \pi} = \int  {d^{4} q\over (2 \pi)^{4}} \;
{1\over q^{2} + m_{\sigma, \pi}^{2}} +
\int {d^{3} q\over (2 \pi)^{3}} \; {1\over \omega_{\sigma, \pi}}
\; n_B (\omega_{\sigma , \pi})  \; ,
\end{equation}
where
\begin{equation}\label{eq28}
\omega_{F}^{2} = \mbox{\boldmath $q$}^{2} + m_{F}^{2} \,
, \; \; \;
\omega_{\sigma,\pi}^2=\mbox{\boldmath $q$}^2+m_{\sigma,\pi}^{2} \, ,
\end{equation}
\begin{equation}\label{eq29}
n_{F}(\omega) = {1\over e^{\beta(\omega - \mu)} + 1} +
{1\over e^{\beta(\omega + \mu)} + 1}  \, ,
\end{equation}
\begin{equation}\label{eq30}
n_{B}(\omega) = {1\over e^{\beta \omega} - 1}     \, .
\end{equation}
The infinite part in (\ref{eq26}) and (\ref{eq27}) is the usual
 $T = \mu = 0$  tadpole
that is absorbed in the tadpole cancellation at the tree level.
Therefore, we can write  our equations
by retaining the $T$- and $\mu$-dependent
pieces only.
Equations (\ref{eq250}) and (\ref{eq251}) finally read
\newpage
\begin{eqnarray}\label{eq32}
\sigma^{2} & = &
f_{\pi}^{2} - {8 g^{2}\over \lambda}     N_{c}
\: \int {d^{3} q\over (2 \pi)^{3}}
\: {1\over 2\omega_F} \;  n_F (\omega_F)
 \nonumber  \\
 &   &
- 3 \: \int  {d^{3} q\over (2 \pi)^{3}}
\: {1\over \omega_{\sigma}} \; n_{B} (\omega_{\sigma})
- 3 \: \int {d^{3} q\over (2 \pi)^{3}}
\: {1\over \omega_{\pi}} \; n_{B} (\omega_{\pi}) \, ,
\end{eqnarray}
\begin{equation}\label{eq320}
s=
\lambda \: \int  {d^{3} q\over (2 \pi)^{3}}
\: {1\over \omega_{\sigma}} \; n_{B} (\omega_{\sigma})
+ 3\lambda \: \int {d^{3} q\over (2 \pi)^{3}}
\: {1\over \omega_{\pi}} \; n_{B} (\omega_{\pi}) \, ,
\end{equation}
where the right-hand sides depend on
$\sigma$ and $s$ through the masses.
These equations have been derived from the
effective potential
 (\ref{eq16})
in which the loop corrections
$\Omega_I$
have been neglected.
This approximation corresponds to the
leading order in the
$1/N$ expansion, where $N$ is the
number of scalar fields \cite{mey}.
In our case, $N=4$.

A straightforward approach to solving
 (\ref{eq32}) and
 (\ref{eq320})
leads to problems with a complex effective potential
\cite{mey}.
 It may easily be seen that
a direct use of  (\ref{eq11})
 leads necessarily to complex solutions.
 From  (\ref{eq32}),
 (\ref{eq320}) with
 (\ref{eq11}) one finds
\begin{equation}
m_{\pi}^2=
 - 8 g^{2}  N_{c}
\: \int {d^{3} q\over (2 \pi)^{3}}
\: {1\over 2\omega_F} \;  n_F (\omega_F)
-2 \lambda \: \int  {d^{3} q\over (2 \pi)^{3}}
\: {1\over \omega_{\sigma}} \; n_{B} (\omega_{\sigma}).
\label{eq321}
\end{equation}
This means that $m_{\pi}^2$ is either negative
or complex.
In both cases it implies
a complex $\sigma$.
In the following sections we show
how a consistent inclusion
of one-loop self- energy corrections removes this problem.

\section{Effective meson masses} \label{pimass}

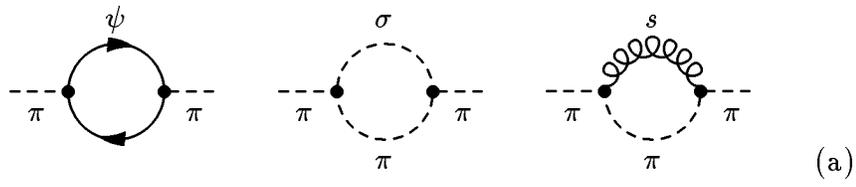
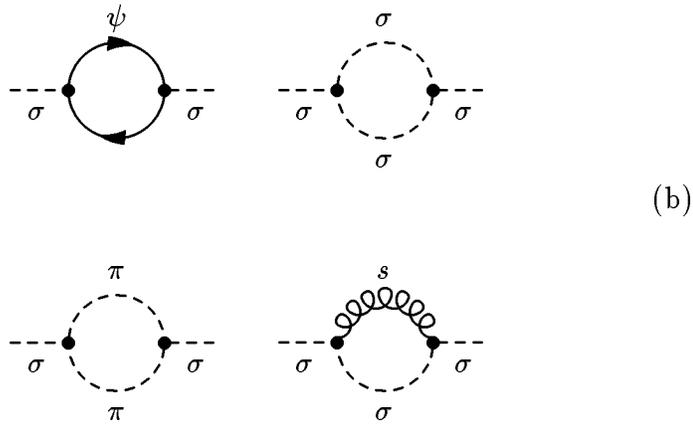
\begin{figure}[p]
\caption{One-loop self-energy diagrams contributing to
         (a) pion    and (b) sigma masses.}
\label{fig2}
\end{figure}

At first sight it seems that the $T, \mu$ dependence of $\sigma$
and $s$ is
not consistent with the Goldstone theorem since
\begin{equation}\label{eq33}
m_0^{2} + s(\beta,\mu) +\lambda \sigma^{2} (\beta, \mu) \neq 0 \; .
\end{equation}
However, the $m_{\pi}^{2}$ must also include the $T$- and $\mu$-
dependent pieces coming from the one-loop (and higher) order
self-energy diagrams (Fig. \ref{fig2}a):
\begin{equation}\label{eq34}
m_{\pi}^{2} = m_0^{2} +
s(\beta,\mu)
 + \lambda \sigma^2(\beta,\mu)
 +\Pi_{\pi}(\beta,\mu) \; ,
\end{equation}
where (at one-loop order)
\begin{eqnarray}\label{eq35}
\Pi_{\pi} & = &
  - 8 N_{c} g^{2} \; {1\over \beta} \sum \int
{d^{3} p\over (2 \pi)^{3}} \; {1\over p^{2} + m_{F}^{2}}
- 4 \: g'^{2} \; {1\over \beta} \sum \int {d^3 k\over (2 \pi)^3}
\; {1\over k^2 + m_{\sigma}^2} \; {1\over k^{2} + m_{\pi}^{2}}
 \nonumber  \\
 &  &
-{1\over \beta} \sum \int {d^3 k\over (2 \pi)^{3}}
\: {1\over k^{2} + m_{\pi}^{2}}(-2\lambda) \; ,
\end{eqnarray}
with $p$ and $k$ defined as in (\ref{eq19}) and (\ref{eq20}),
respectively.
The $s$-field appears in the
Lagrangian
  without a kinetic term and its
 propagator is simply
$1/m_s^2=-2\lambda$.

The combined contribution of the
$\sigma$-tadpoles
 (\ref{eq250}) and the $s$-tadpoles
 (\ref{eq251})
is given by
\begin{eqnarray}\label{eq36}
m_0^{2} + s+\lambda \sigma^2
 & = &
 \underbrace{m_0^{2} + \lambda
f_{\pi}^{2}}_{= 0}
+8\: g^{2} N_{c} \; {1\over \beta} \sum \int
{d^{3} p\over (2\pi)^{3}} \; {1\over p^{2} + m_{F}^{2}}
 \nonumber  \\
&  &
- 2 \: \lambda \; {1\over \beta} \sum \int {d^{3} k\over (2 \pi)^{3}}
\; {1\over k^{2} + m_{\sigma}^{2}} \, .
\end{eqnarray}
Combining (\ref{eq35}) and (\ref{eq36}), we see that
the fermion parts cancel immediately, whereas the boson parts cancel
 provided the relation
\begin{equation}\label{eq37}
m_{\sigma}^2- m_{\pi}^2=2\lambda\sigma^2 \, ,
\end{equation}
which holds trivially at $T = \mu = 0$, also holds
  at
 $T, \mu \leq T_{c}, \mu_{c}$.
Thus, the consistency with the Goldstone theorem requires that
the relation
\begin{equation}\label{eq37a}
  m_{\sigma}^{2} = 2 \lambda \sigma^{2}
\end{equation}
should also hold
at $T, \mu \leq T_{c}, \mu_{c}$.
Indeed, we shall shortly
demonstrate that it works
at one-loop level.

We do not agree here with Larsen \cite{lar} who obtained
$m_{\pi} \neq 0$ even in the symmetry-broken phase,
thus violating the Goldstone theorem.
The reason is that he assumed
$m_{\pi} \neq 0$
in the propagators in the one-loop self-energy diagrams.
If
$m_{\pi} =0$ at the tree level, then
the mass corrections would be of order
$\lambda$ or higher.
This would in turn  yield self-energy corrections
of  order $\lambda^2$, which would then also require
the inclusion of two-loop diagrams.
In other words, it is not consistent with one-loop
calculations to include
corrections to the tree-level masses
in the propagator.

Similarly as for the pion mass, for the sigma mass we have
\begin{equation}\label{eq38}
m_{\sigma}^{2} = m_0^{2} + s(\beta,\mu)
+3 \lambda \sigma^{2}(\beta,\mu) +
\Pi_{\sigma}(\beta,\mu) ,
\end{equation}
where the self-energy of the $\sigma$
 particle (Fig. \ref{fig2}b) is given by
\begin{eqnarray}\label{eq39}
\Pi_\sigma & = &  - 8 g^2 N_c \; {1\over \beta}
\sum \int {d^3p\over (2\pi)^3} \left( {1\over p^2 + m_F^2} -
{2 m_F^2\over (p^{2} + m_{F}^{2})^{2}} \right)
 \nonumber  \\
 &  &
-3\cdot 6 g'^2 \; {1\over \beta} \sum \int {d^3 k\over (2 \pi)^3}
\; {1\over (k^2 + m_{\sigma}^2)^2}
-2 \cdot 3 g'^2 \; {1\over \beta} \sum \int
{d^{3} k\over (2 \pi)^{3}}\; {1\over (k^{2} + m_{\pi}^{2})^2}
 \nonumber  \\
 &  &
- {1\over \beta} \sum \int {d^3 k\over (2 \pi)^3}
\; {1\over k^2 + m_{\sigma}^2} (-2\lambda)
\: .
\end{eqnarray}
This, together with the tadpole part
\begin{eqnarray}\label{eq40}
m_0^{2} + s+3 \lambda \sigma^{2}
 & = &
\underbrace{m_0^{2} + 3 \lambda
f_{\pi}^{2}}_{2\lambda f_{\pi}^2}
+24\: g^{2} N_{c}  \; {1\over \beta} \sum \int
{d^{3} p\over (2\pi)^{3}} \; {1\over p^{2} + m_{F}^{2}}
 \nonumber  \\
 &   &
- 8  \lambda \: {1\over \beta} \sum \int {d^3 k\over (2 \pi)^3}
\: {1\over k^{2} + m_{\sigma}^{2}}
-6 \lambda \: {1\over \beta} \sum \int {d^{3} k\over
(2 \pi)^{3}} \: {1\over k^{2} + m_{\pi}^{2}}  \; ,
\end{eqnarray}
 gives
\begin{eqnarray}\label{eq41}
m_{\sigma}^{2}
  & = &
 2 \lambda f_{\pi}^{2} + 2\cdot 8 \: g^{2}
N_{c}  {1\over \beta}\sum \int \frac{d^3p}{(2\pi)^3}\;{1\over p^2+
m_{F}^{2}}
\nonumber  \\
  &   &
- 2\cdot 3 \: \lambda {1\over \beta}\sum \int {d^3k\over (2\pi)^3}
\; {1\over k^{2} + m_{\sigma}^{2}}
- 2 \cdot 3 \: \lambda {1\over\beta}\sum \int {d^3k\over (2\pi)^3}
\; \frac{1}{(k^2 + m_{\pi}^2)}  + ... \; ,
\end{eqnarray}
where ...
denotes terms of higher order in $\lambda$ and $g^{2}$.
Combining this equation
with (\ref{eq250}), we recover
(\ref{eq37a}).
Thus, the thermal corrections at one-loop level do not
alter the tree-level
equation for the effective sigma mass.

\section{Chiral restoration transition}  \label{chiral}

From
the analysis in the preceding section
we see that the $s$-dependence
of the $m_{\sigma}$ and
$m_{\pi}$ is removed owing to the one-loop
self-energy corrections.
Equations
(\ref{eq32}), (\ref{eq320}) in which we put
\begin{eqnarray}\label{eq43}
m_{\pi} & = & 0 \; ,
\nonumber  \\
m_{\sigma}^2 & = & 2 \lambda \sigma^{2} \; ,
\nonumber  \\
m_{F} & = & g \sigma \; ,
\end{eqnarray}
decouple and
we are able to determine the
 $T$ and $\mu$ dependence of the chiral condensate by solving
 (\ref{eq32}) only.
If one compares our approach with the standard one
where  the quartic interaction is kept,
one may easily check that the self-energy diagrams
with internal $s$-lines
plus the contribution of $s$-tadpoles
precisely equals the contribution
of the loops with the quartic vertex.
In this way, the saddle-point method
\cite{fre,mey}
becomes equivalent to the standard
approach.

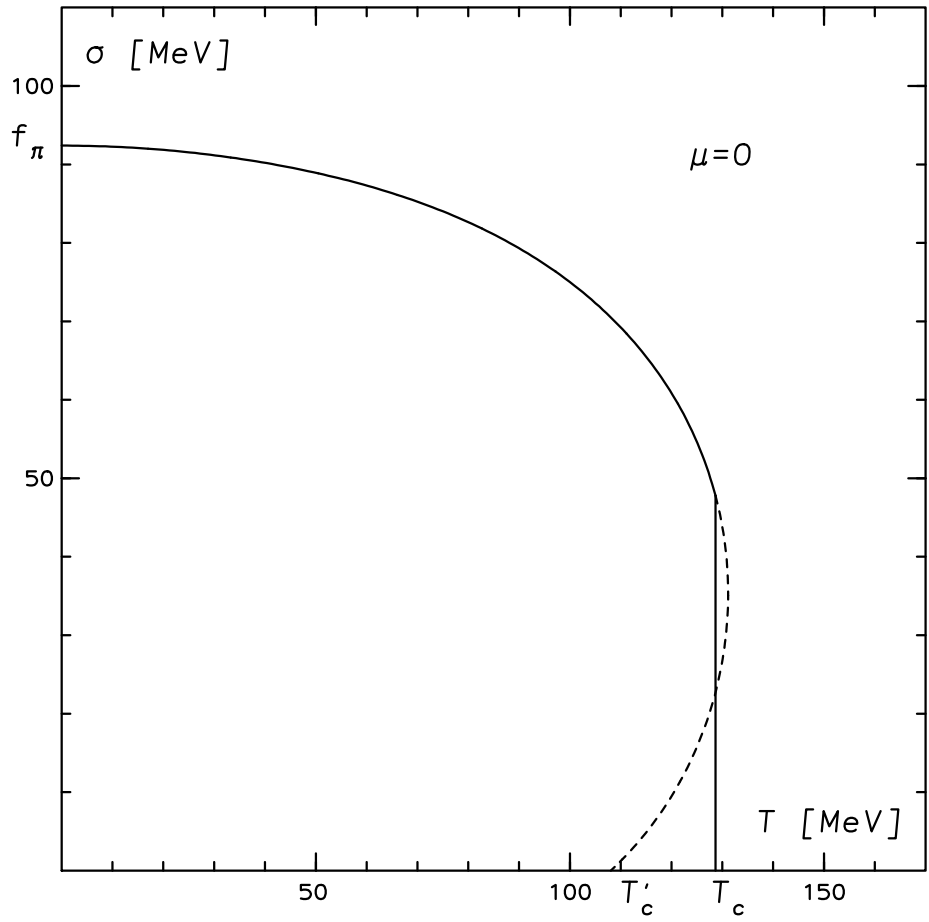
\begin{figure}[p]
\caption{Chiral condensate as a function of temperature
at
$\mu =0$.
The dashed line corresponds to a physically unstable solution.
 $T_{c}$ is
the temperature of the first-order phase transition.}
\label{fig3}
\end{figure}

For $\mu=0$, we find the solution numerically as a function of $T$
depicted in Fig. {\ref{fig3}.
As input parameters we choose the constituent quark mass
$m_F=340$ MeV and the sigma mass $m_{\sigma}=1$ GeV.
 The solution indicates a first-order
phase transition.
The point where the curve crosses the $T$ axis
is not the point of the actual phase transition.
We  refer to it  as ``critical" point,
having in mind that it would be a critical
point if the phase transition were second order.
The actual transition takes place at the point
 where the two
minima of the effective potential at $\sigma=0$
and $\sigma (T_c)$ are leveled.
Hence, the transition temperature $T_c$ is determined by requiring
\begin{equation}\label{eq46}
\Omega (\sigma, T_c) = \Omega (0, T_c) \, .
\end{equation}

The solution is particularly simple in the neighborhood of the
``critical" point $T_{c}'$ since in this case we can
expand the integrands around
$\sigma=0$
and perform the
integrals analytically \cite{kap}.
We find
\begin{eqnarray}\label{eq44}
\sigma^{2}& =& f_{\pi}^{2} - \left[ {2 g^{2}\over \lambda} \;
N_{c} \; {T^{2}\over 6} + {T^{2}\over 2}
\right]
+{3\sqrt{2\lambda} \over 4\pi} \sigma T
\nonumber
\\
 & &
+\sigma^2 \left[ \left({\gamma \over 2} -{1 \over 4}\right)
\left({3\lambda \over 2\pi^2} -{2g^4N_c \over \lambda\pi^2}\right)
+{3\lambda \over 4\pi^2} \ln
{\sqrt{2\lambda}\sigma \over 4\pi T}
-{g^4 N_c \over \lambda\pi^2} \ln
{g\sigma \over \pi T} \right]
+{\cal O}(\sigma^3) \, ,
\end{eqnarray}
where $\gamma=0.5772...$ is Euler's constant.
At the ``critical" temperature,  $\sigma (T_c') = 0$ , which
gives
\begin{equation}\label{eq45}
\frac{T_c'^2}{f_{\pi}^2} = \frac{6\lambda}{2g^2
 N_c + 3\lambda}.
\end{equation}
If we had taken the Nambu-Jona-Lasigno relation
$m_{\sigma}=2m_F$ \cite{nam},
which is exact in the  \\
$N_c\rightarrow\infty$
limit \cite{fis},
we would have obtained $T_c'=f_{\pi}$.
We stress again that
$T_c'$  only approximates
the actual transition temperature $T_c$
to be determined from the condition (\ref{eq46}).

If we put $N_c=1$ in
  (\ref{eq45}) our result agrees with that of Anand et al.
\cite{ana}
who considered a similar model for nuclear matter in
 Walecka's mean-field approach.
Our result (\ref{eq45}) also agrees with Bochkarev and Kapusta
\cite{boc} if  we put $N_c=0$.
However,
some of the existing calculations that include fermions
\cite{ram,con,cle,lur}
disagree with our result for the following reasons.
Ram Mohan \cite{ram},
and subsequently
Contreras and Loewe \cite{con}
did not properly account for the contribution
of antifermions.
As a consequence, they obtained fewer
(by a factor of two, apart from color)
fermionic degrees of freedom,
which in turn yielded a larger estimate for the $T_c$
\cite{lur}.
The reason for disagreement with
Cleymans, Koci\'c and Scadron \cite{cle}
is twofold.
First, in their
 calculation of
the sigma mass
they did not include
all the relevant  self-energy diagrams
(Fig. \ref{fig2}).
Second,
the sign of the fermionic contribution to
the sigma mass is wrong.
This led again to a larger $T_c$ in terms of
$f_{\pi}$.
The agreement of their estimate of
the critical temperature
with the estimates based on
a single  meson-loop diagram \cite{bai}
and  on the Nambu-Jona-Lasigno
model \cite{bil} is only accidental.

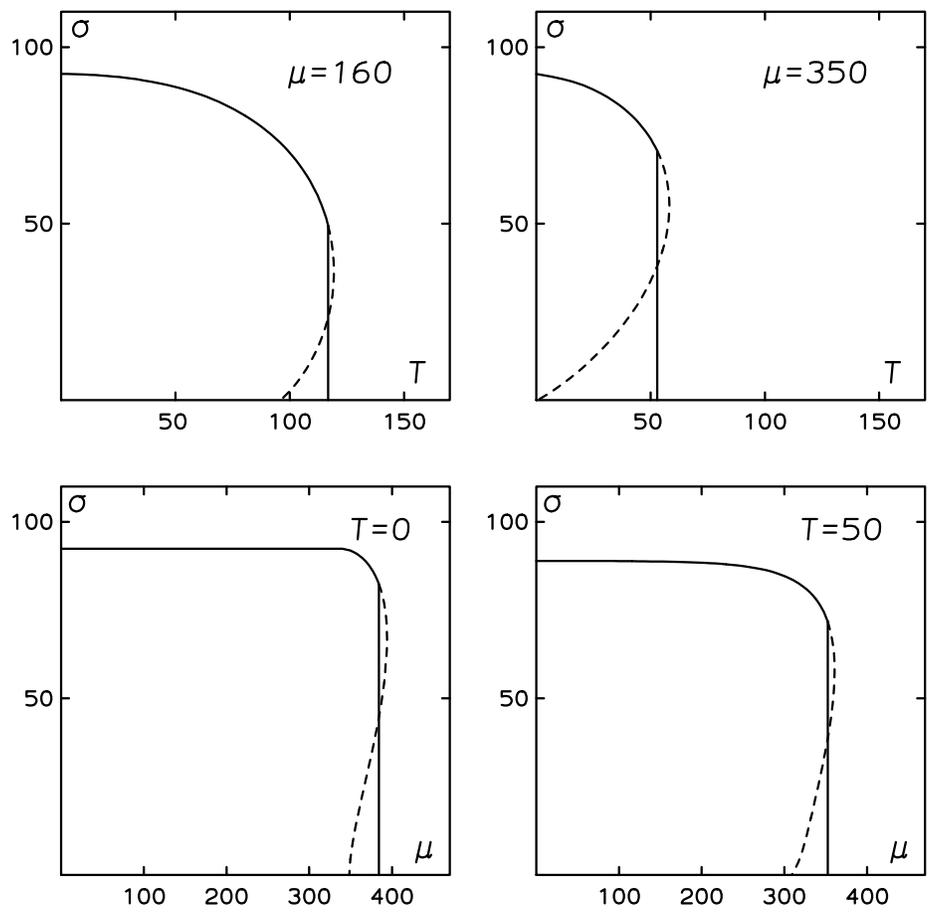
\begin{figure}[p]
\caption{
Chiral condensate at nonzero baryon density.
$\sigma$, $T$ and $\mu$ are in MeV.}
\label{fig4}
\end{figure}

The calculations at nonzero chemical potential are similar.
We present our results in \\
 Fig. \ref{fig4}.
The chiral condensate
$\sigma$ is plotted as a function of temperature
for fixed
 $\mu$ (160  and
 350 MeV; upper two plots)
 and as a function of chemical potential for fixed
 $T$ (0  and
 50 MeV;
 lower two plots).
 The transition remains first order and
 the critical temperature decreases with $\mu$
 as expected.

 In Fig. \ref{fig5} we plot the phase diagram
 of nuclear matter and compare it with the
 thermal parametrization of recent heavy-ion collision
 data.
 The baryonic chemical potential is related to the
 quark chemical potential as
 $\mu_B=3\mu$.
 The phase boundary
 between the chirally symmetric and broken phases
 (solid line)
 appears to be very close to
 the expected phase boundary
 between hadron resonance gas and quark-gluon plasma
 with a bag constant \cite{har}.
 The diagram indicates that the chiral transition
 might slightly precede the deconfinement.
 The points with error bars show the freeze-out
 values of $T$ and $\mu$ deduced from AGS \cite{bra1,let,raf} and SPS
\cite{bra2} data with flow.

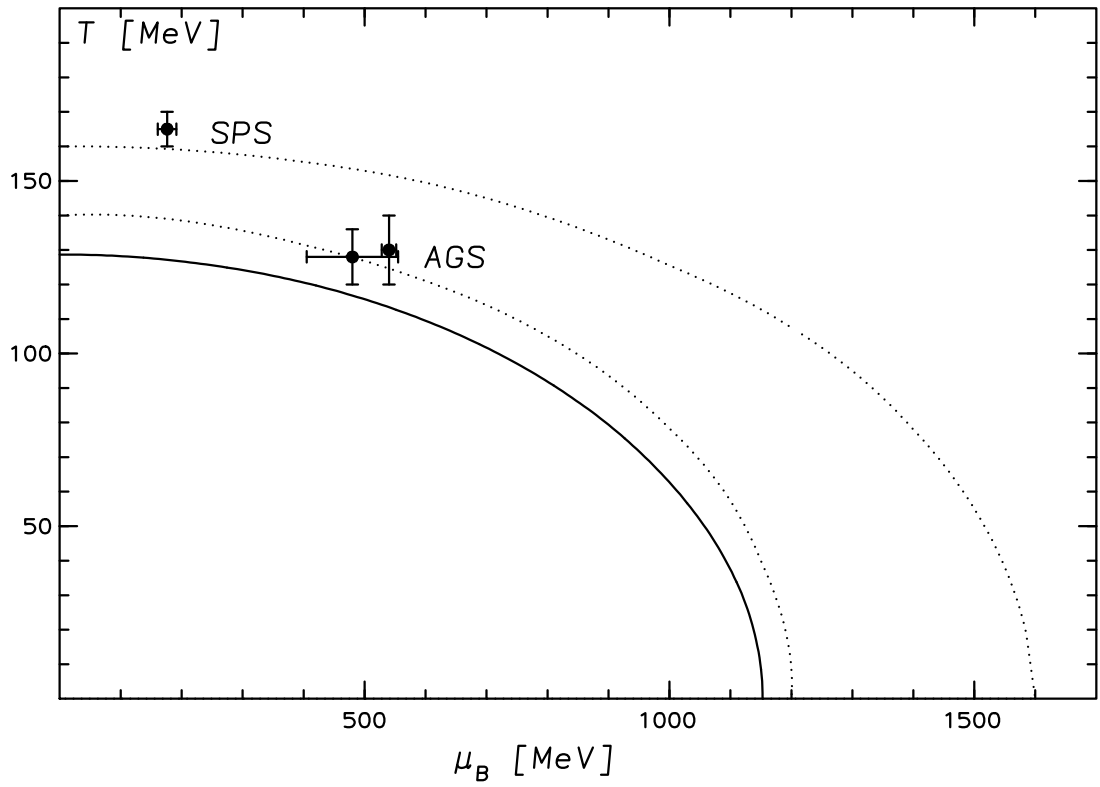
\begin{figure}[p]
\caption{Phase diagram as a plot of
 temperature versus baryon chemical
    potential.
The solid line separates the chirally broken (inside)
    and chirally symmetric (outside) phases.
The
expected hadron gas --
 quark-gluon plasma boundary is located between
the two dashed lines.}
\label{fig5}
\end{figure}

In Fig. \ref{fig6} the phase diagram is represented in terms
of temperature and baryon density.
The solid line separating the chirally symmetric and
broken phases shows how the transition temperature
depends on baryon density.

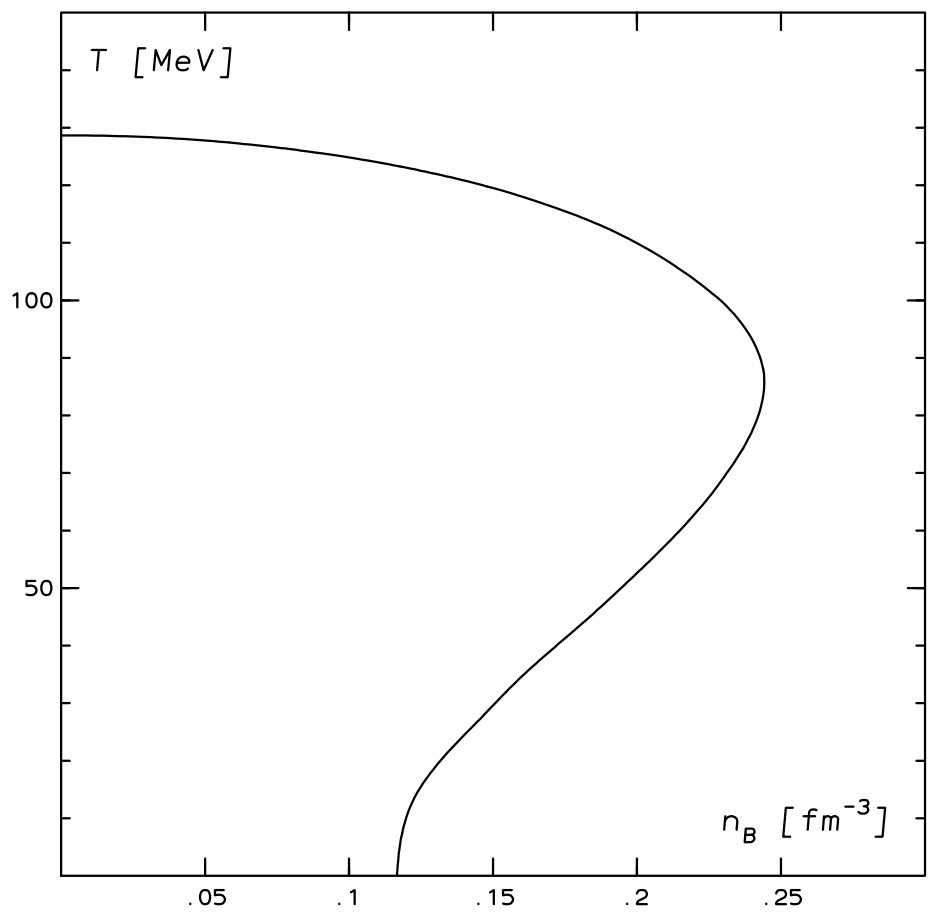
\begin{figure}[p]
\caption{Phase diagram as a plot of
 temperature versus baryon density.}
\label{fig6}
\end{figure}

Similarly to the chiral condensate, the effective masses will
have a discontinuity at $T_c$.
Above the critical line in Fig. \ref{fig5}
the symmetry is restored and the
$\sigma$-tadpoles  vanish from the theory.
The $T$ and $\mu$ dependence of
the meson mass
$m_M=m_{\pi} = m_{\sigma}$ is determined by the proper self-energy
diagrams and $s$-tadpoles:
\begin{equation}\label{eq47}
m_M^{2} = m_0^{2} +
s(\beta,\mu)
 +\Pi_{\pi}(\beta,\mu) \;.
\end{equation}
Here $s(\beta,\mu)$ and
 $\Pi_{\pi}(\beta,\mu)$
 may  be calculated
  at lowest order in $\lambda$ and $g^2$ using
(\ref{eq251})
and (\ref{eq35}) in which the propagator masses
and $g'$ are set to zero.
In particular, for $\mu=0$ and above $T_c$ we find the
following
 expression  for the meson mass
\begin{equation}\label{eq48}
m_M^{2} = \left(\frac{\lambda}{2}
+\frac{N_c}{3}g^2\right)(T^2-T_c'^2).
\end{equation}

\section{Conclusion} \label{concl}
We have shown that the usual mean-field pattern of symmetry breaking
and restoration gives a consistent picture in the $\sigma$-model at the
one-loop order.
We have shown that the saddle-point method
\cite{fre,mey}
is equivalent to
a standard approach
\cite{ram,ana,boc}
if the self-energy loop corrections are included.
The phase transition is predicted to be
first order, in agreement with the analysis of Pisarski and Wilczek
\cite{pis}.
The chiral phase boundary in a ($T,\mu_B$) plot
(Fig. \ref{fig5})
is close to the phase boundary between
  hadron  gas  and
 quark-gluon plasma.
The thermal parametrization of
 existing experimental data compared with
 the chiral phase diagram indicates that
 the nuclear matter produced in heavy-ion collisions
 is close to or slightly above the chiral-phase-transition
 line.
 It is therefore conceivable that the present and future
 heavy-ion experiments may observe effects of the chiral
 transition.
 A more sophisticated analysis of the data is needed
 in order to observe
 possible signatures
 near the critical density,
 such as
 a rapid change of the meson mass and width
 or abnormal production ratios of charged to neutral
 pions \cite{har}.

\section*{Acknowledgement}

NB would like to thank J. Cleymans,
    D. Luri\'e
and M. Scadron for useful discussions.

\newpage

\begin{figure}[p]
\label{fig7}
\end{figure}
\end{document}